# Double Indirect Interlayer Exciton in a MoSe$_2$/WSe$_2$ van der Waals Heterostructure

*Aubrey T. Hanbicki,*[‡*] *Hsun-Jen Chuang,*[‡a] *Matthew R. Rosenberger,*[b] *C. Stephen Hellberg, Saujan V. Sivaram,*[b] *Kathleen M. McCreary, I.I. Mazin, and Berend T. Jonker*

*Materials Science & Technology Division, Naval Research Laboratory,* Washington, DC 20375, USA

[a] Postdoctoral associate at the Naval Research Laboratory through the American Society for Engineering Education
[b] Postdoctoral associate at the Naval Research Laboratory through the National Research Council
[‡] These authors contributed equally to this research
[*] Correspondence and requests for materials should be addressed to A.H. (email: hanbicki@nrl.navy.mil)

ABSTRACT
An emerging class of semiconductor heterostructures involves stacking discrete monolayers such as the transition metal dichalcogenides (TMDs) to form van der Waals heterostructures. In these structures, it is possible to create interlayer excitons (ILEs), spatially indirect, bound electron-hole pairs with the electron in one TMD layer and the hole in an adjacent layer. We are able to clearly resolve two distinct emission peaks separated by 24 meV from an ILE in a MoSe$_2$/WSe$_2$ heterostructure fabricated using state-of-the-art preparation techniques. These peaks have nearly equal intensity, indicating they are of common character, and have *opposite* circular polarizations when excited with circularly polarized light. *Ab initio* calculations successfully account for these observations – they show that both emission features originate from excitonic transitions that are indirect in momentum space, are split by spin-orbit coupling, and that including interlayer hybridization is essential in correctly describing the ILE transition. Although well separated in momentum space, we find that in real space the electron has significant weight in both the MoSe$_2$ and WSe$_2$ layers, contrary to the commonly assumed model. This is a significant consideration for understanding the static and dynamic properties of TMD heterostructures.

**Keywords**
Transition metal dichalcogenides; 2D heterostructure; MoSe$_2$; WSe$_2$; two-dimensional materials; valley polarization; photoluminescence;

Tailoring semiconductor heterostructures for specific functionalities has led to varied opto-electronic devices including solar cells, photodetectors, light-emitting diodes and lasers. An emerging class of heterostructures involves stacking discrete monolayers such as the transition metal dichalcogenides (TMDs)[1,2] to form so-called van der Waals heterostructures (vdWHs)[3,4]. vdWHs offer novel functionalities, making them promising hosts for future devices. One unique new property is the formation of an interlayer exciton (ILE), a spatially indirect, bound electron-hole pair with the electron in one TMD layer and the hole in an adjacent layer[5–14].

$MoSe_2/WSe_2$ is a bilayer heterostructure composed of isoelectronic Mo and W diselenide monolayers. It has a type II band alignment with a *spatially indirect* minimal excitation gap, with the top of the valence band formed predominantly by W states and the bottom of the conduction band by Mo states[15–18]. ILE emission in this heterostructure has recently been observed[8–13], indicating significant dipole transitions between layers. The reported photoluminescence (PL) energy of the ILE emission is in the range of 1.35 – 1.4 eV. Due to the type II band alignment of the heterostructure[15–18], this energy is well separated from the emission energies of the constituent $MoSe_2$ (1.55 eV)[19] and $WSe_2$ (1.65 eV)[20] monolayers. As in the isolated monolayers, the heterostructure violates inversion symmetry, resulting in spin-orbit splitting of the bands. While interlayer excitons have been reported in systems such as $WSe_2/MoS_2$ [5], $MoS_2/WS_2$ [6,7], and $MoSe_2/MoS_2$ [14], we confine our discussion and comparison to the $MoSe_2/WSe_2$ [8–13] system because the lattice matching and ordering of the conduction band splitting in other systems could produce fundamentally different results.

Two groups have reported a splitting of the ILE in $MoSe_2/WSe_2$ at low temperature, although the splitting was not well resolved[8,13]. The origin of this splitting and indeed of the ILE itself has not been clarified. One group reports the splitting to be on the order of 25 meV and suggests the two peaks originate from the bright and dark excitons at the K-point, with both transitions direct in momentum-space[8]. The 25 meV splitting in emission energy agrees well with the calculated *ab initio* spin-orbit (SO) splitting of the $MoSe_2$ conduction band at the K-point[21]. Another group deconvolves their data into two peaks separated by almost 40 meV[13]. They propose one of the peaks corresponds to a transition indirect in real space yet direct in momentum space, while the second feature is indirect in both real and momentum space. Both scenarios are inconsistent with the similar intensity observed for the two ILE peaks.

In this work, by using advanced preparation techniques, we fabricate a vdWH in which we are able to resolve the ILE splitting clearly for the first time, enabling us to elucidate the nature of the ILE and the origin of these features. The split emission features exhibit nearly equal intensity and opposite polarizations that vary in a non-monotonic fashion with excitation energy. Based on considerations of their relative intensities, polarizations, and *ab initio* calculations, we conclude that both transitions are indirect in momentum space, in contrast with previous interpretations. The valence band maxima (hole states) occur at the K, K' points in the Brillouin zone, while the conduction band minima (electron states) occur at the Q, Q' points. Both bands exhibit splittings due to SO effects. Furthermore, although the electron in the interlayer exciton is commonly thought to reside entirely in the Mo-layer, we find instead that it has significant weight in both layers at Q. In contrast, the electron states reside entirely in the Mo-layer at the K-point. We find that including interlayer hybridization is essential to theoretically determine the ILE character. The hybridized electron eigenstates are superpositions of both spin states, and both spin-orbit split bands are optically bright, decaying optically with holes at the K point with opposite polarizations. The Q-K transition is suppressed in momentum space relative to a putative direct transition at K, because it is indirect and requires either a phonon or defect scattering to conserve momentum. But since the relevant wave function at Q has comparable weight in both layers, it has significant overlap in real space. The lowest-energy direct transition at K is suppressed by the point symmetry (see Supplementary information). This scenario is qualitatively different from previous models and accounts for the roughly equivalent emission intensity of both peaks, and is consistent with the raw data previously reported for this heterostructure[8,13]. Our layer- and spin-resolved band structure calculations provide insight into the origin of the ILE, and suggest ways to tailor the indirect / direct momentum space character of one or both transitions.

## RESULTS

**MoSe$_2$ on WSe$_2$ heterostructures**. We prepared a number of MoSe$_2$ on WSe$_2$ heterostructures, and a schematic of the resulting system is shown in **Fig. 1a**. Typical results are summarized here, and a more detailed account is presented in the Supplementary Information (SI). The individual monolayer components were synthesized using chemical vapor deposition (CVD) and transferred with a dry transfer technique described in the Methods section and illustrated in Fig. SI-1. Two samples discussed in the main text are shown in the optical micrograph of **Fig. 1b**. In this image, there are two monolayer (ML) WSe$_2$ triangles on top of hexagonal boron nitride (hBN) with a larger ML MoSe$_2$ flake draped over them. The edges of both WSe$_2$ triangles in **Fig. 1b** are aligned within < 3° of the edges of the MoSe$_2$. Hereafter, we will refer to the MoSe$_2$/WSe$_2$ overlap regions as S1 and S2, as labeled in the figure. A third sample (S3) was also fabricated with a misalignment of ~28°. In each case, the structure is capped with a second layer of hBN (not shown in **Fig. 1b** for clarity). Atomic force microscopy (AFM) images from all of our samples are presented in Figs. SI-2 and SI-3.

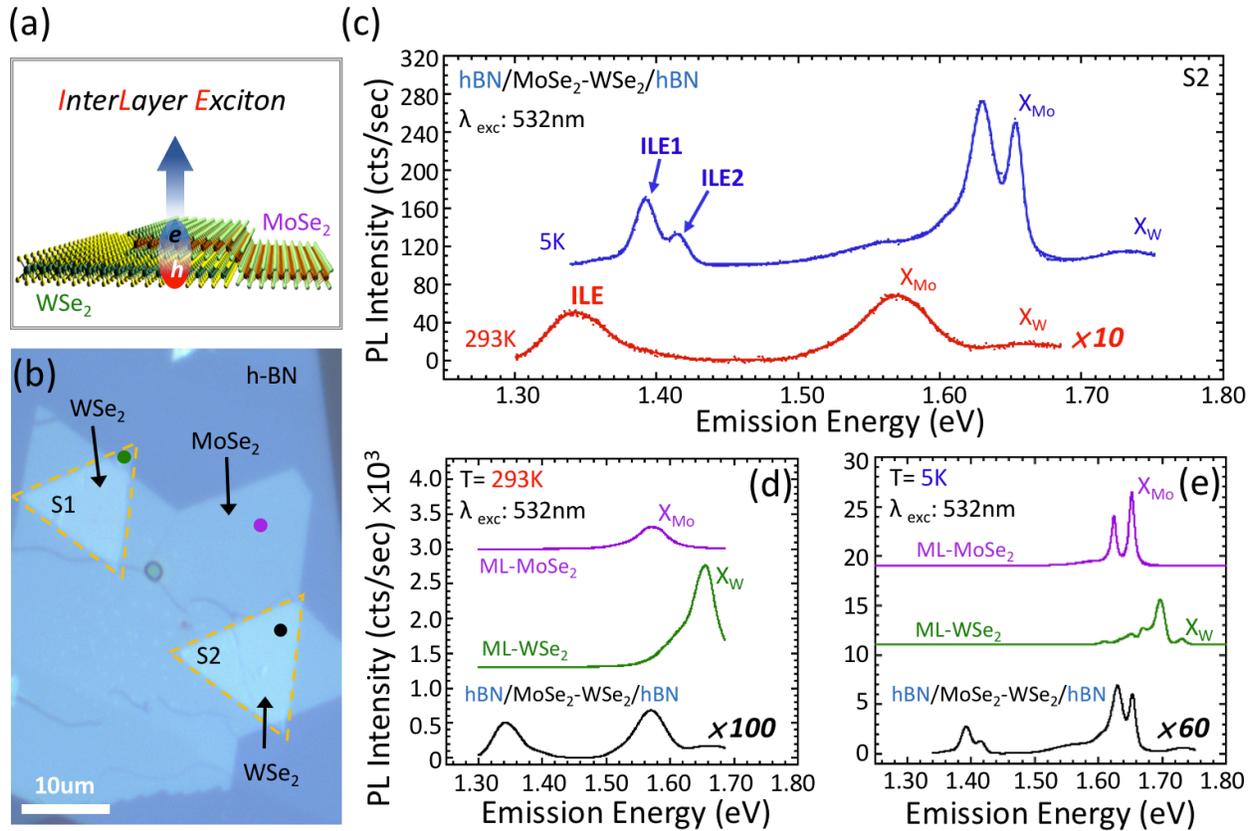

**Figure 1 | Characterization of the interlayer exciton.** (a) Schematic depiction of the MoSe$_2$/WSe$_2$ heterostructure. (b) Optical micrograph of two vdWH samples, S1 and S2, before capping with the top hBN layer for ease of visualization. The yellow dotted line indicates the outline of CVD ML-WSe$_2$ and the larger star-like shaped layer is the CVD ML-MoSe$_2$. The colored dots indicate the position where PL data were acquired. (c) PL of the interlayer exciton from sample S2 at room temperature (red) and 5 K (blue). A comparison of the PL from the interlayer exciton, monolayer MoSe$_2$, and monolayer WSe$_2$ regions are shown at (d) 293 K and (e) 5 K. Spectra in (c-e) are offset for clarity and scaling factors are indicated as necessary. The ground state exciton emission features from the individual MoSe$_2$ and WSe$_2$ layers are labeled $X_{Mo}$ and $X_W$, respectively.

To reduce inhomogeneity and increase intimate contact between layers, we incorporated several advances in sample preparation. First, we have encapsulated the entire structure within hBN, which significantly reduces the inhomogeneous contributions to PL linewidths by providing surface protection as well as substrate flatness[22,23]. Even in encapsulated samples, interlayer imperfections persist[24]. Therefore, we also flattened areas of the sample using an AFM as a squeegee, as described in the Methods section and shown in Figs. SI-2 and SI-3. We are thus able to remove nearly all the residual material between the two TMD monolayers in a select area of the overlap region, resulting in an intimate and reproducible contact.

A summary of the PL from various spots on our sample is shown in **Fig. 1c-e** using an excitation energy of 2.33 eV (532 nm). The physical location where each spectrum was collected is indicated by dots on **Fig. 1b** and color coded with the spectra. As expected, reference PL spectra from the individual $MoSe_2$ and $WSe_2$ layers exhibit strong peaks at 1.57 eV and 1.65 eV, respectively, at room temperature (**Fig. 1d**) and 1.65 eV and 1.71 eV at 5 K (**Fig. 1e**). In the encapsulated and AFM flattened overlap regions S1 and S2, the $WSe_2$ and $MoSe_2$ emission is strongly quenched as can be clearly seen in these figures. This is expected and has been attributed to the ultra-fast charge separation enabled by the close proximity of these monolayers[25]. Our PL and reflectivity lines are somewhat broader than those reported in single layers of encapsulated TMDs[22,23], albeit much narrower than in unencapsulated heterostructures. Intrinsic broadening of linewidths in heterostructures has been reported elsewhere and attributed to the fundamental optical processes in heterostructures[26]. Further broadening could be due to the relative quality of our samples as well as the lengthy procedure in ambient required to assemble the structure[24]. We also note that the dielectric environment of a heterostructure will be somewhat different than that of individual layers sandwiched between hBN. Further characterization of all our samples, including Raman spectroscopy, is presented in the SI section SI-2 and corroborates the interacting nature and reproducibility of the $MoSe_2/WSe_2$ vdWHs.

**An interlayer exciton emerges.** A new emission feature emerges at 1.35 eV at room temperature in the overlap regions (**Fig. 1c**). This feature has commonly been associated with the ILE[8–13]. The intensity of the ILE PL is slightly different for both samples, and the spatial variation of the ILE peak intensity is mapped in Fig SI-2 for both S1 and S2. In the flattened region, the ILE emission is mostly uniform (Fig. SI-2). The ILE is also observed for a sample that was flattened but not encapsulated, sample S0. It is not observed in S3, the intentionally misaligned heterostructure, consistent with recent reports[11]. Further characterization of S0-S3 are presented in section SI-2.

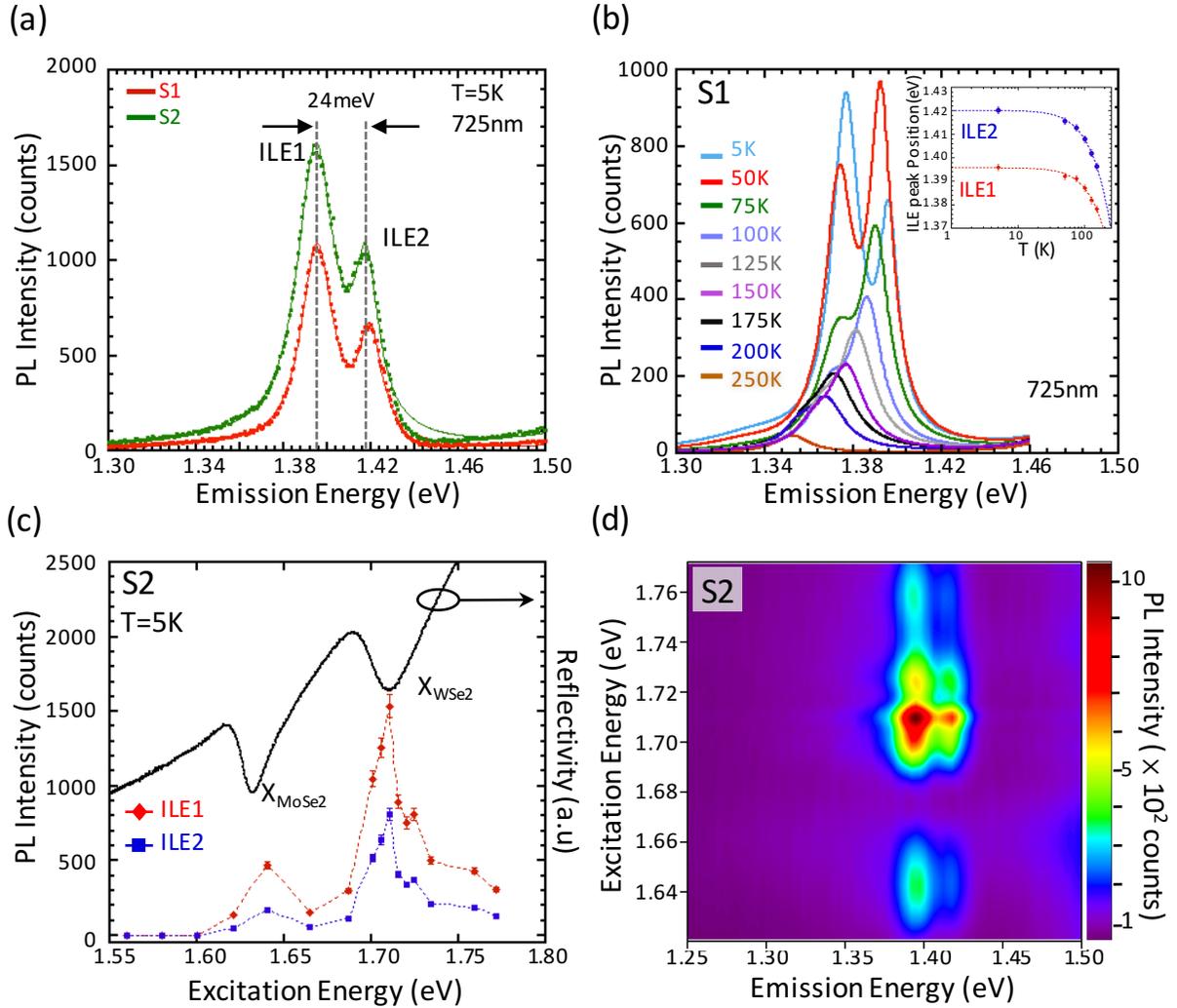

**Figure 2 | Emission from two ILEs:** (a) Interlayer exciton PL at low temperature showing two well-resolved peaks, ILE1 and ILE2. An excitation energy of 1.710 eV (725 nm) was used, and spectra from both S1 (red) and S2 (green) are shown. (b) Temperature dependent PL measurement of S1. The inset shows the peak shift for each emission line and follows a standard semiconductor behavior. (c) PL Peak intensity of ILE1 (red points) and ILE2 (blue points) at different excitation energies from S2. The black line is the differential reflectivity measurement of S2. (d) PLE heat map of S2 showing splitting of the ILE and intensity maxima at resonant energies.

When the sample is cooled to 5 K, the ILE peak exhibits a well-resolved splitting with distinct peaks at 1.396 eV (ILE1) and 1.420 eV (ILE2), as shown in **Fig. 2a** for samples S1 and S2. A similar plot for all of the samples is presented in Fig. SI-8. The linewidths of these peaks are 20 meV (ILE1) and 13 meV (ILE2), significantly narrower than the widths of 45 meV and 30 meV reported by Rivera *et al.*[8]. The high resolution of our spectra enable us to accurately determine the splitting to be 24 ±1 meV, similar to the splitting inferred by Rivera *et al.*[8]. This splitting is seen over the entire flattened overlap area. The ILE splitting is unlikely to be caused by a charged exciton because the trion binding energy in $MoSe_2$ is greater than 30 meV[27–29]. The temperature dependence of the ILE features us shown in **Fig. 2b**. A full discussion of the temperature dependence and power dependence of these peaks is provided in sections SI-3 and SI-4. In brief, the temperature dependence of the peak positions follows a standard semiconductor behavior, as shown by the inset to **Fig. 2b** and in Fig. SI-13, providing good evidence of intrinsic behavior from a uniform, intimate contact between our layers. The evolution of the relative strength of the two peaks with both increasing temperature and excitation power is presented in Figs. SI-9 to SI-14.

The data in **Figs. 2a,b** were taken with an excitation wavelength of 1.710 eV (725 nm). The reason for this choice is clear from **Figs. 2c,d**. **Fig. 2c** shows the PL peak intensity for ILE1 (red points) and ILE2 (blue points) at 5 K as a function of excitation energy. The differential reflectivity (**Fig. 2c** - solid black line) shows a strong correlation of optical absorption with the maxima in the ILE PL intensity at 1.71 eV. A heat map of these data is shown in **Fig. 2d**, confirming this correlation. We did not observe any signature of a charged exciton transition in differential reflectivity. This indicates a small oscillator strength and suggests both monolayers have a low level of doping[29].

**Circular Polarization.** A very striking behavior is observed upon excitation with circularly polarized light. We find that ILE1 and ILE2 both exhibit significant polarization, as reported previously[9,13], but contrary to these reports, we find the polarizations are of *opposite* sign and exhibit a non-monotonic dependence upon excitation energy, as shown in **Fig. 3**. Most single layer TMDs are semiconductors with a direct gap[30] at the K-points and are well known for their potential as valleytronic materials because they have two inequivalent, but related K-points in the Brillouin zone, K and K' [31–33]. By symmetry, the valence band maxima at K and K' have opposite spin states, giving these materials unique optical selection rules[31–34]. Using circularly polarized light, it is possible to selectively populate and interrogate the different valleys, K or K', and valuable information on the nature of the bands can be derived from studies of polarization-resolved emission.

The spectra in **Fig. 3** were obtained with a circularly polarized excitation source of energy 1.710 eV (725 nm). The excitation has positive helicity (σ+) and we analyze the PL for positive and negative helicity (σ−). Polarization is defined as $P = [ I(σ+) – I(σ−) ] / [ I(σ+) + I(σ−) ]$, where $I(σ ±)$ is the emission intensity analyzed for positive (negative) helicity. It is clear from the raw spectra shown in **Fig. 3a,b** for vdWH samples S1 and S2 that the ILE1 and ILE2 emission peaks have opposite circular polarizations. These spectra can be well fit with two

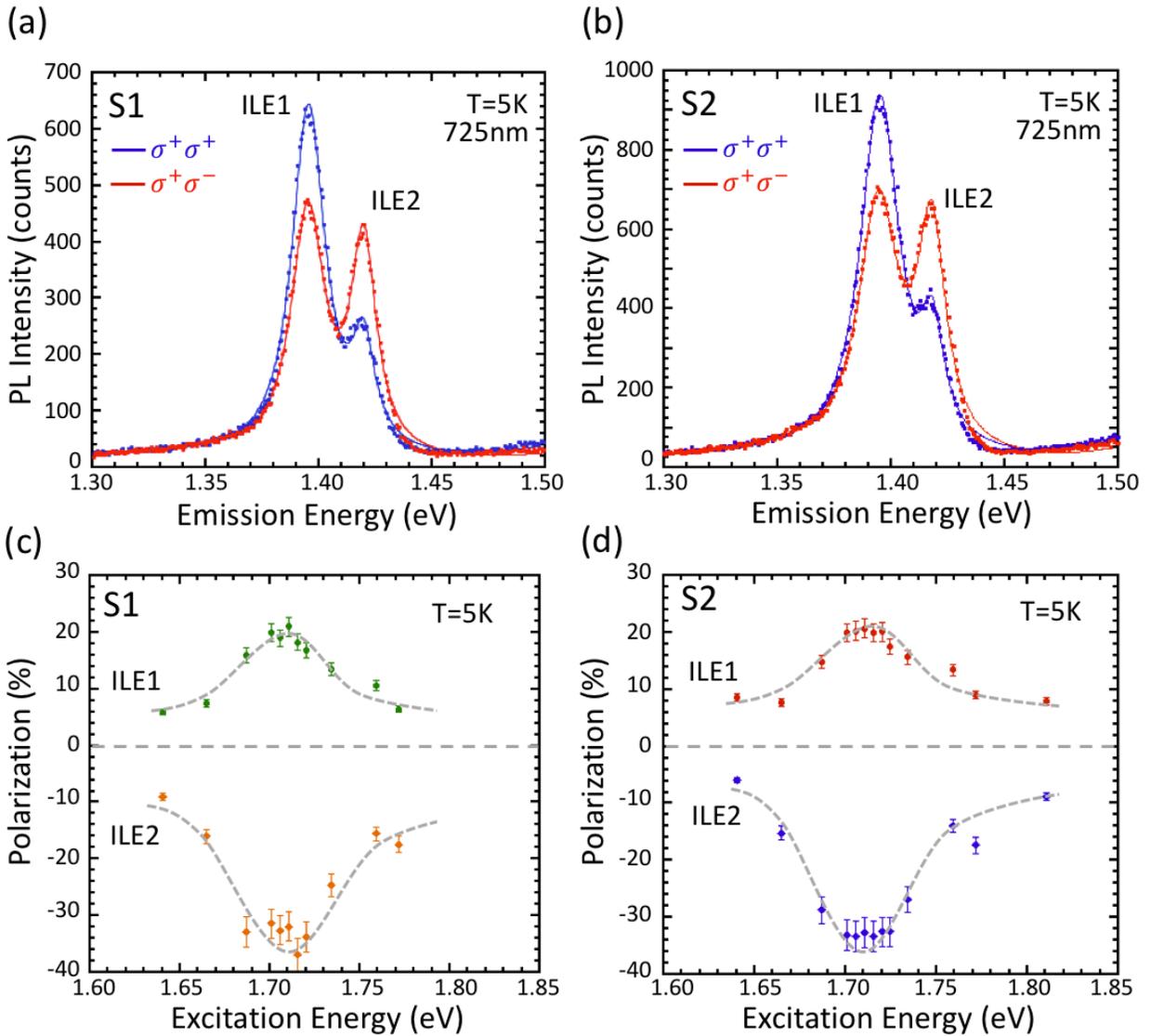

**Figure 3 | Opposite circular polarization from each ILE.** PL measurements using a circularly polarized excitation at 1.710 eV (725 nm) and analyzing for same (blue) and opposite (red) helicity for (a) S1 and (b) S2 at 5 K. A summary of the polarization of each peak as a function of excitation energy is shown for (c) S1 and (d) S2. The gray dotted guiding line is a guide for the eye.

Lorentzians, as illustrated in Fig. SI-6, to quantitatively determine the polarization of each feature. The polarization of ILE1 and ILE2 as a function of excitation energy is shown in **Fig. 3c,d** for vdWH samples S1 and S2, respectively. Significantly, the peaks have opposite polarizations for all excitation energies, and the magnitude of the polarization for each feature exhibits a pronounced non-monotonic behavior with a strong peak at 1.71 eV, corresponding to the absorption feature in the WSe$_2$ (**Fig. 2c**). This behavior persists to 120 K, as shown in Fig. SI-11. Prior studies on the MoSe$_2$/WSe$_2$ heterostructure have shown only a positive polarization for the ILE[9,13]; because separate peaks were never clearly resolved, the behavior shown in **Fig. 3** was not visible. Polarizations of opposite sign were noted in a system where monolayer WSe$_2$ was subjected to a large magnetic field[35], but the reason or mechanism for the opposite handedness was not determined. The fact that the magnitude of polarization is nearly twice as large for ILE2 suggests that the opposite signs are not symmetry-defined, but have a quantitative nature.

**DISCUSSION**

In order to understand this polarization behavior and the detailed nature of the ILE excitons, we have computed the band structure of the MoSe$_2$/WSe$_2$ vdWH using density functional theory (DFT). The results are shown in **Fig. 4a**, where the color coding indicates the layer from which the states are derived and the arrows indicate the spin orientation. Details of the calculations are given in the Methods section. The valence band maximum (VBM) lies at K and K' and is localized entirely within the WSe$_2$ layer; the corresponding W-derived states have the quantum numbers $L_z = 2$, $S_z = 1/2$, so the spin at the VBM, indicated by arrows in **Fig. 4a**, is parallel to $z$, the direction perpendicular to the layers. The lowest conduction band at K is a pure Mo-derived state, with $L_z = -1$, $S_z = 1/2$. However, the minimal gap is indirect, and occurs at a point between Γ and K. Isoenergetic surfaces (**Fig. 4b**) indicate that the absolute conduction band minimum (CBM) is located at the Q-point.

The electron density of the conduction band at the band edges is shown in **Fig. 4c**. From this figure it is clear the CBM at Q is strongly hybridized, with significant contributions from both the MoSe$_2$ and WSe$_2$ layers – note the common isosurfaces on both the Mo and W atoms. In contrast, both the conduction band and valence band edge states at K exhibit negligible hybridization. There are two reasons for this: first, the K-point has a distinct point symmetry that additionally suppresses hybridization, while Q is a general point, and second, the bare W and Mo bands come closer to each other at Q than at K. The CBM has considerable contribution from the $WSe_2$ layer, which facilitates optical decay between electrons at Q and holes at K, the VBM. The hybridization between layers mixes bands with opposite spins resulting in the tilting of the spins in the CB away from K. While mixing at Q is evidence of interlayer overlap at this point in the Brillouin zone, one can prove that intralayer localization at K is dictated by symmetry and does not indicate a total absence of overlap. This is addressed in the Supplementary Information, section SI-6. The SO splitting of the VBM at K is large, while that in the conduction band at Q is much smaller.

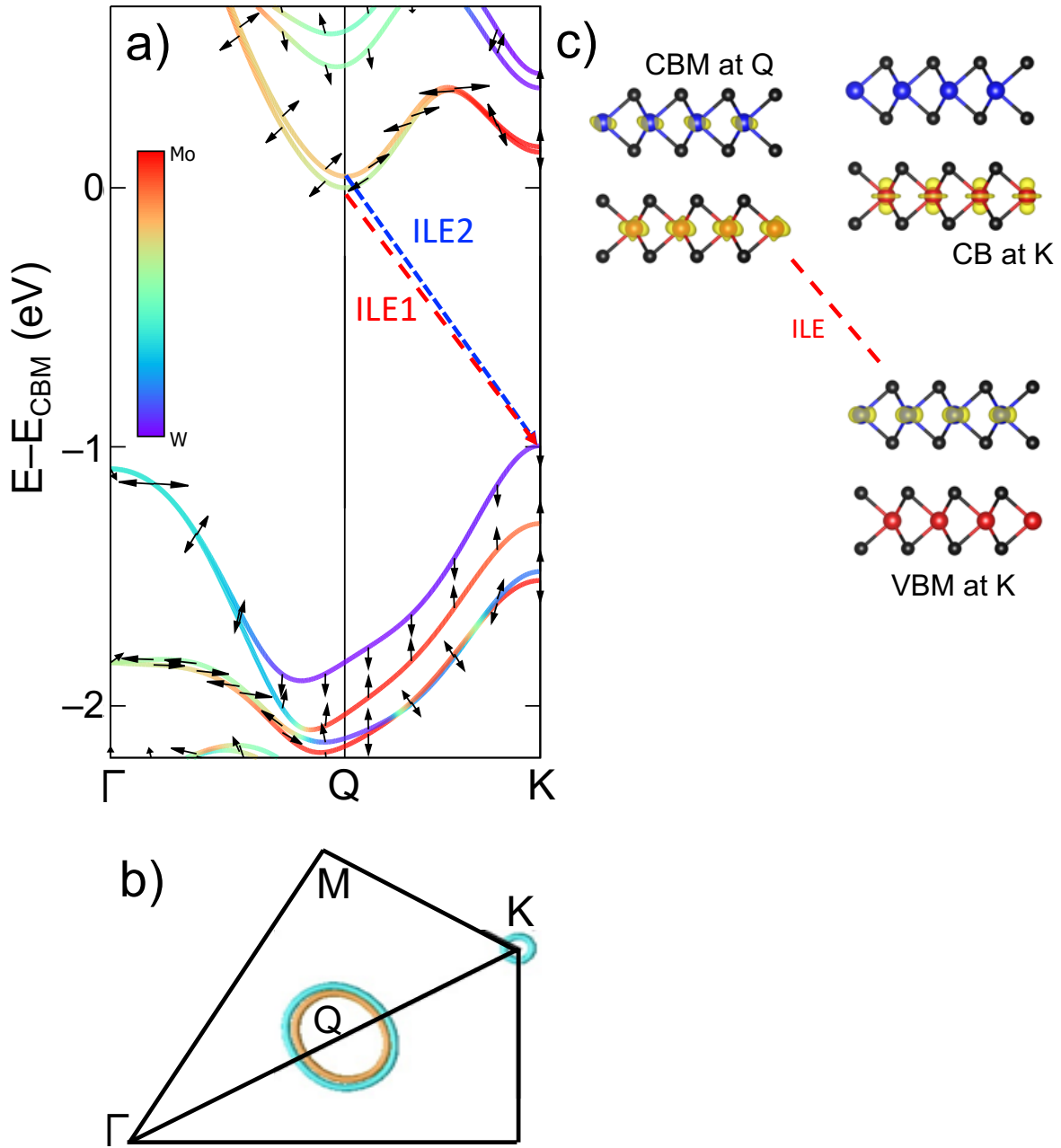

**Figure 4 | Band structure of the MoSe$_2$/WSe$_2$ heterostructure.** (a) Band structure of the MoSe$_2$/WSe$_2$ heterostructure. The conduction band minimum lies at the Q point. The color indicates the layer from which the states are derived: red is pure MoSe$_2$, blue is pure WSe$_2$. The spin direction (integrated over the entire cell) is in the *y-z* plane (*x* is defined by Γ-K). The two lowest energy transitions, ILE1 and ILE2 are indicated with dashed lines. (b) The isoenergetic surfaces in the conduction band for the irreducible wedge of the Brillouin zone. The lowest two states occur at the Q-point (SO-split), while the third lowest state appears at K. (c) Graphical representation of the electron density isosurfaces at the three points of the band structure indicated. The Mo are red, W are blue, Se are black, and the electron density isosurfaces are yellow. Higher resolution figures are shown in the Supplementary information.

The fact that symmetry forbids hybridization between W and Mo at the K point implies that the corresponding transition will be dark. As shown in section SI-6, numerical calculations of the direct optical transition find that both SO-split transitions have a vanishing optical weight in the vicinity of K and cannot be associated with either of our observed excitons, contrary to some earlier proposals[8,13]. Meanwhile, the CBM at the Q point has considerable weight on W; the challenge is to calculate the dipole matrix elements for circularly polarized photons inside the W atomic sphere. Based on the wave function character in density functional calculations (see Supplementary Information, section SI-6) we estimate that the two indirect Q-K transitions are both allowed and have similar weight. Numerically, recombination between the CBM at Q and VBM at K emits predominantly positive helicity light, as observed for ILE1, while recombination between the higher SO split CB state at Q and the VBM at K emits mostly negative helicity photons, as observed for ILE2. Thus, we conclude with confidence that (1) the observed ILE occurs between the upper valence state at K and the two lowest, SO-split conduction states at Q and (2) while symmetry selection rules are relaxed for such double-indirect excitons, the character of the states at Q is such that the ILE splits into predominantly positive- and predominantly negative-helicity peaks, a novel feature that should be investigated for potential applications. Note that these are semi-quantitative considerations. For a complete and quantitative theory, calculations beyond static DFT are highly desirable.

In summary, we have fabricated high quality $MoSe_2/WSe_2$ heterostructures, and for the first time observe a well-resolved splitting of the interlayer exciton photoluminescence arising from the SO splitting of the conduction band. The two split features exhibit significant (20-35%) and opposite circular polarizations when excited with circularly polarized light, with maximum polarization occurring when excited resonantly at the $WSe_2$ absorption peak. By analyzing the individual properties of the two peaks, comparing with first principles theory, and using general symmetry considerations, we are able to exclude the proposal that one or both of the two excitons are due to direct transitions at the K points. Instead, we find the two interlayer excitons are fully indirect in momentum space, and partially indirect in real space. Contrary to the accepted model, the electron in the ILEs have significant weight in both layers due to interlayer hybridization. The hybridization of the electrons between the layers tilts the electrons' spins, and both ILEs are optically bright with opposite polarizations. Our results have important implications on the utility of these systems for future electronic and valleytronic applications.

# METHODS:

## SAMPLE PREPARATION:
### CVD growth:
Monolayer $MoSe_2$ and $WSe_2$ are synthesized via atmospheric pressure CVD using solid precursors in a 2-inch quartz tube furnace. Silicon wafers with 275 nm thermally grown oxide (Silicon Valley Microelectronics, Inc.) and single-side polished c-plane sapphire (University Wafer) are used as the target substrates for $MoSe_2$ and $WSe_2$, respectively. Prior to growth, the substrates are cleaned by ultrasonication, piranha etching and oxygen plasma exposure. Both TMD syntheses use a water-soluble seeding promoter, perylene-3,4,9,10-tetracarboxylic acid tetrapotassium salt (PTAS), that is dropcast onto a clean $SiO_2$ substrate immediately before growth. A clean $SiO_2$ substrate or sapphire substrate is downstream from the PTAS substrate for $MoSe_2$ or $WSe_2$, respectively. The substrates are loaded face down on a quartz boat directly above the solid precursor (50 mg of $MoO_2$ (Sigma Aldrich, 99%) or 1000 mg $WO_3$ (Alfa Aesar, 99.998%). The precursor and substrates are then moved to the center of the tube furnace. An additional quartz boat containing Se (Alfa Aesar, 99.999%) is placed upstream near the edge of the furnace. The tube is evacuated to ~100 mTorr and repeatedly filled with UHP Ar and $H_2$. For $MoSe_2$, monolayer synthesis occurs at 750 °C, whereas $WSe_2$ requires a higher temperature of 950 °C.

### Sample transfer:
The hBN encapsulated heterostructure is prepared via a water-assisted pick up, dry transfer method and is fully detailed in the Supplementary Information, section SI-1, which includes a schematic of the transfer process. The full structure from top to bottom is 5 nm-hBN / ML-$MoSe_2$ / ML-$WSe_2$ / 20 nm-hBN / 275 nm-$SiO_2$ / Si substrate for S1 and S2. In brief, the stack was picked up in reverse order with freshly prepared PDMS then transferred to the substrate.

### PDMS preparation:
Polydimethylsiloxane (PDMS) is made from a commercially available SYLGARD 184 Silicone Elastomer Kit. To make the PDMS mixture, we mix the two components thoroughly (Silicone Elastomer and curing agent) with the weight ratio of 10:1 followed by a debubbling process under rough vacuum. This mixture is spin coated on a silicon wafer with a spin rate of 350 rpm for 30 seconds, then cured at 80˚C for 30 minutes on a hotplate. The resultant PDMS is easily peeled off the silicon wafer for use.

### AFM flattening:
This technique more thoroughly removes residual material from between layers than the commonly used technique of annealing[11,36] and has a significantly smaller thermal budget since no heating is required. AFM flattening was performed on a Park Systems NX-10 AFM. The AFM cantilever used for flattening was an NCHR (Nanosensors) with a nominal spring constant of 42 N/m. The scan size for AFM flattening varied from 6 μm to 15 μm, depending on the desired size of the flattened region. The scan rate was typically 1 Hz corresponding to a scan speed as high as 30 μm/s. The scan line density was typically 10 nm/line or less in order to maintain sufficient overlap between lines, which caused contaminants to be squeezed out of the flattened area rather than accumulating between scan lines. The required normal force to achieve good flattening depended on sample/tip specific parameters, including the tip radius and hBN thickness. Generally, thicker hBN required larger normal force. To determine the appropriate normal force, the tip was first engaged with the minimum possible force and then the force was increased while observing the topography. The force was increased until bubbles and wrinkles disappeared from the topography.

## SAMPLE CHARACTERIZATION:
Temperature dependent PL, Raman and differential reflectivity spectra are acquired under vacuum conditions (pressure~$1 \times 10^{-5}$ Torr) in a Janis ST-500 Microscopy Cryostat using a commercial Horiba LabRam HR Evolution confocal spectrometer. Beam steering mirrors control the laser position in the *x-y* sample plane and enable both single spot and scanned area acquisition. Excitation sources include a tunable continuous-wave (CW) Ti: Sapphire laser (Spectra-Physics) as well as various single wavelength CW lasers for PL and Raman, and a white light source (Energetiq - LDLS) for differential reflectivity. To enable comparison between the various lasers we have used only cw sources[37]. A 50X objective (NA=0.35) is used to focus the laser to a spot of ~2 μm diameter. A quarter wave plate (Thorlabs superachromatic) is used to circularly polarize the laser excitations. The resulting photoluminescence is collected and directed through the same quarter wave plate and a subsequent rotatable linear polarizer to analyze the circularly polarized emission components. We obtain the same polarization when the sample is excited with negative helicity light, and the emitted circular polarization is 0% when the sample is excited with linearly polarized light.

THEORY:
Most of the calculations presented here were performed using the generalized-gradient approximation[38] with the DFT-D3(BJ) van der Waals correction[39,40] and projector augmented wave functions as implemented in the Vienna *ab initio* simulation package (VASP)[41–43]. A plane wave cutoff of 450 eV and a 4x4 Γ–centered *k*-point mesh was used. Atomic positions were relaxed until residual forces were less than 0.5 meV/Å. The optimal lattice constant of the bilayer was found to be 3.28 Å. 20 Å of vacuum was used between periodic images normal to the layers. The potentials included the following orbitals in the valence: Mo ($4d^45s^2$), W ($5d^46s^2$), and Se ($4s^24p^4$). The all-electron WIEN2K package was used to confirm the accuracy of the band structure computed with VASP and to compute optical matrix elements[44].

**Acknowledgements**
This research was performed while H.-J.C. held an American Society for Engineering Education fellowship and M.R.R. and S.V.S. held a National Research Council fellowship at NRL. This work was supported by core programs at NRL and the NRL Nanoscience Institute, and by the Air Force Office of Scientific Research under contract number AOARD 14IOA018-134141. This work was also supported in part by a grant of computer time from the DoD High Performance Computing Modernization Program at the U.S. Army Research Laboratory Supercomputing Resource Center.

**Author Contributions**
B.T.J., H.-J.C and A.T.H. conceived of and designed the experiment. S.V.S. and K.M.M. fabricated samples and H.-J.C. assembled the heterostructures. A.T.H., H.-J.C. and M.R.R. performed the experiments. A.T.H., H.-J.C. and M.R.R. analyzed the data. C.S.H and I.I.M performed the first principles calculations. All authors discussed the results and contributed to the writing of the manuscript.

**Additional Information**
**Supplementary information** accompanies this paper.

**Competing financial interests:** The authors declare no competing financial interests.

Supplementary Information

# Double Indirect Interlayer Exciton in a MoSe$_2$/WSe$_2$ van der Waals Heterostructure

*Aubrey T. Hanbicki,[†] Hsun-Jen Chuang,[†a] Matthew R. Rosenberger,[b] C. Stephen Hellberg,*

*Saujan V. Sivaram,[b] Kathleen M. McCreary, I.I. Mazin, and Berend T. Jonker*

*Materials Science & Technology Division, Naval Research Laboratory,*
Washington, DC 20375, USA

[a] Postdoctoral associate at the Naval Research Laboratory through the American Society for Engineering Education
[b] Postdoctoral associate at the Naval Research Laboratory through the National Research Council
[†] These authors contributed equally to this research

**SI-1. Transfer method**

The hBN encapsulated heterostructure is prepared via a water-assisted pick up, dry transfer method. This transfer method is based on previously reported transfer techniques and is shown schematically in **Fig. SI-1**.[1–3] Monolayer (ML) material of MoSe$_2$ and WSe$_2$ were grown via a CVD method detailed in the methods section and single crystal hBN is commercially obtained (MOMENTIVE® PT110). Heterostructures were constructed in the following steps. First, the bottom multilayer-hBN is exfoliated onto a 275 nm-SiO$_2$/Si substrate. Next, PDMS is prepared as detailed in the methods section. The top thin multilayer-hBN is then exfoliated directly onto the freshly made PDMS as a carrier flake. The first monolayer (TMD1) is picked up with the PDMS/hBN (**Fig. SI-1a**). The second monolayer (TMD2) is carefully aligned and subsequently picked up with the PDMS/hBN/TMD1 stack (**Fig. SI-1b**). This final heterostructure (PDMS/hBN/TMD1/TMD2) is transferred onto the hBN/SiO$_2$/Si base (**Fig. SI-1c**).

**SI-2. Detailed Characterization of Multiple Samples**

To confirm the reproducibility of our results, we have fabricated multiple samples, S0-S3. A summary of the samples is presented in **Table SI-1**. All samples have the same stacking order except S0 which does not have a top hBN capping layer. The bottom hBN is typically 20 nm thick or greater and the top hBN is ~10 nm thick. A portion of the overlap area is then flattened using our AFM squeegee technique.[4] The relative alignment between the monolayers, the qualitative interlayer exciton PL signal before and after flattening, and the area flattened on each sample is detailed in **Table SI-1**. Data from S1 and S2 are featured in the main text.

**Figure SI-2a** shows the optical image of each completed heterostructure as well as an AFM image and a schematic diagram detailing the relative placement and orientation of each layer. Also indicated is the area that was flattened (black box) and the position where room temperature PL was collected. A noticeable rounded bubble appears in the AFM image of S1 and S2 after the AFM flattening process and is due to contaminants trapped between the thick bottom hBN and the $SiO_2$/Si substrate. It does not affect the heterojunction. The PL spectra are presented in **Fig. SI-2b** and were taken with an excitation energy of 2.33 eV (532 nm).

A detailed AFM image of sample S3 is shown in **Fig. SI-3**. These images are typical of all the samples. In this figure, the step heights of CVD ML-$WSe_2$ and ML-$MoSe_2$ are 7.5 and 6.9 Å respectively. The top and bottom hBN are 9.5 nm and 19.5 nm respectively. Surface roughness (RMS) measured on the entire hBN/$MoSe_2$/$WSe_2$/hBN stack in the flattened region is less than 100 pm.

**Table SI-1.**
Summary of 4 samples used in this study. The misalignment angle is determined from the AFM image in **Fig. SI-2**.

| Sample | Structure | $MoSe_2$/$WSe_2$ Angle θ | ILE signal before flattening | ILE signal after flattening | Approximate flattened area |
|---|---|---|---|---|---|
| S0 | $MoSe_2$/$WSe_2$/hBN | 57.1° | weak | strong | 6μm x 6μm |
| S1 | hBN/$MoSe_2$/$WSe_2$/hBN | 57.4° | weak | strong | 15μmx15μm |
| S2 | hBN/$MoSe_2$/$WSe_2$/hBN | 58.5° | weak | strong | 12μmx12μm |
| S3 | hBN/$MoSe_2$/$WSe_2$/hBN | 28° | No | No | 8μm x 15μm |

Room temperature Raman spectra from the flattened region in all of the samples is presented in **Fig. SI-4**. In these spectra, not only is there a clear in-plane $E^1_{2g}$ feature but the $B^1_{2g}$ feature expected from both bilayer $MoSe_2$ and $WSe_2$ is also visible. This is a strong indication of an interaction between the monolayers. Note that these features are even apparent in S3 where the monolayers are misaligned by ~30°. Despite the mismatch in angle, these monolayers also appear to be interacting although the $B^1_{2g}$ intensity does depend on the alignment angle.

Low temperature (5 K) photoluminescence spectra from the overlap region, S2, as well as adjacent, encapsulated, isolated $MoSe_2$ and $WSe_2$ are presented in **Fig. SI-5**. These spectra were taken over an energy range where the emission from the individual monolayers as well as the ILE are all present. Components of both $MoSe_2$ and $WSe_2$ are visible in the heterostructure, however, they are significantly quenched relative to the emission from the adjacent monolayers indicating a strong interaction between the layers. Further, the ILE intensity is nearly as large as the $MoSe_2$ emission and significantly higher than the $WSe_2$ emission.

## SI-3. Comparison of S1 and S2

To better understand the reproducibility of our results we present a comparison of results from S1 and S2 in **Fig. SI-7–11**. To fit our PL spectra, we use Lorentzian peaks for ILE1 and ILE2, include a low energy defect peak at low temperature and power, and include a small background contribution from the tail of the laser. In **Fig. SI-6**, an example is shown of a PL spectrum fit to these components. In this example, the PL was measured from S1 at 5 K with an excitation energy of 1.71 eV (725nm). The line widths here, which are typical, are 18.5 meV and 12.6 meV for ILE1 and ILE2, respectively.

Using this fitting procedure, we fit the PL spectra from our samples and have plotted the polarization (**Fig. SI-7**) and PL intensity (**Fig. SI-8**) as a function of excitation energy with data from both S1 and S2 on the same plot. It is clear from these plots both samples yield qualitatively and quantitatively similar results. For the polarization data, we also include the data using an excitation energy significantly higher than the emission energy (2.33 eV; 532 nm). For the PLE data, we also include the differential reflectivity (**Fig. SI-8** – black line). Temperature dependent data for the intensity (**Fig. SI-9**), width (**Fig. SI-10**), and polarization (**Fig. SI-11**) of both ILE1 and ILE2 as a function of temperature for samples S1 and S2 are presented in the subsequent figures. The intensity of ILE2 has an interesting feature where it increases when the temperature is raised from 5 K to 50 K, then drops as temperature increases beyond 50 K. The width increases monotonically with increasing temperature for both interlayer excitons, as expected. The polarization of each ILE decreases monotonically and approaches zero around 150 K.

## SI-4. Temperature Dependence

The temperature dependence of the ILE is presented in **Fig. SI-12** for S1 (**Fig. SI-12a,b**) and S2 (**Fig. SI-12c**). From these spectra, the peak position of ILE1 and ILE2 was extracted and plotted as a function of temperature (**Fig. SI-13**). We fit these data in two ways. First we used the traditional Varshni formulation[5] (**Fig. SI-13a**):

$$E_g(0) = E_0 - \alpha T^2 / (T + \beta) \quad (1)$$

Where $E_0$ is the zero-temperature energy and $\alpha$ and $\beta$ are Varshni fit parameters. Another formulation we used to fit these data follows O'Donnell[6] (**Fig. SI-13b**):

$$E_g(0) = E_0 - S\langle\hbar\omega\rangle [\coth(\langle\hbar\omega\rangle/2kT) - 1] \quad (2)$$

Here $E_0$ is the zero-temperature energy, $S$ is a dimensionless coupling constant and $\langle\hbar\omega\rangle$ is an average phonon energy. From these fits, we see the peak position of ILE1 is 1.396 eV and ILE2 is 1.420 eV giving a splitting of 24 meV at T = 0 K. The full results of the fitting using equation

(1) are summarized in **Table SI-2** and equation (2) in **Table SI-3**. Unusual variations in the shift in energy as a function of temperature have been reported and attributed to extrinsic effects,[7] therefore we note that our well-behaved temperature dependence is further evidence of intrinsic behavior from a uniform, intimate contact between our layers.

**Table SI-2.**
Fit parameters for fitting temperature dependent peak position using equation (1).

| Varshni et al | $E_0$ (eV) | $\alpha$ (eV/K) | $\beta$ (K) |
|---|---|---|---|
| Sample1 ILE1 | 1.3957 | 3.4E-4 | 284.81 |
| Sample1 ILE2 | 1.4201 | 3.6E-4 | 190.25 |
| Sample2 ILE1 | 1.3959 | 4.1E-4 | 429.66 |
| Sample2 ILE2 | 1.4195 | 3.0E-4 | 151.07 |

**Table SI-3.**
Fit parameters for fitting temperature dependent peak position using equation (2).

| Odonnell et al. | $E_0$ (eV) | S | $\langle \hbar\omega \rangle$ (eV) |
|---|---|---|---|
| Sample1 ILE1 | 1.3955 | 1.048 | 0.011 |
| Sample1 ILE2 | 1.4200 | 1.320 | 0.009 |
| Sample2 ILE1 | 1.3958 | 0.942 | 0.011 |
| Sample2 ILE2 | 1.4192 | 1.225 | 0.009 |

**SI-5. Power Dependence**

There are two distinct effects in the low temperature PL power dependence (**Fig. SI-14**). First, there is a monotonic blue shift of both ILE1 and ILE2 as the excitation power increases. Because interlayer excitons are aligned permanent dipole moments, as the density increases, there is a repulsive dipole-dipole interaction leading to an increase in energy. This phenomenon has been explored thoroughly in the literature of spatially indirect excitons in gallium arsenide (GaAs) coupled quantum wells.[8,9]

There is also an evolution of the relative strength of the two peaks with increasing excitation power, as shown in **Fig. SI-14b,d**. The following is a likely scenario. At low power, the lowest energy configuration of interlayer excitons, an electron in a lower energy band of MoSe$_2$, would be populated first. Due to phase space filling effects, the interlayer exciton configuration with the electron in the next highest energy band starts to be filled at higher laser power. A similar dynamic in peak intensity is also seen in the temperature dependence and fits the same model.

**SI-6. Theory details**

*Absence of interlayer hybridization at the K-point:*
The conduction band at the K-point is formed by $L = 2$, $L_z = -1$ spherical harmonics on the Mo sites. If we project W sites onto the Mo plane, the W falls right at the center of a Mo triangle. The relative phases of the wave functions at these three Mo sites are defined by the symmetry of the K points; specifically, they change by $i\pi/3$ from site 1 to site 2 to site 3. **Fig. SI-15** illustrates that the $Y_{2,-1}$ harmonics combined with proper phases and re-expanded around the center of the triangle only have $L \geq 3$ components, and do not hybridize with $L = 2$ orbitals. A similar consideration holds for the VBM, formed by $L = 2$, $L_z = 2$ harmonics. These expand into $L = 1$ states and again do not hybridize with $L = 2$. Thus, no matter how large the overlap of the wave function may be, the states in question at the K point remain layer-pure. This is, of course, not true for an arbitrary point in the Brillouin zone, in particular at Q. This is made more clear with electron density isosurfaces shown in **Fig. SI-16**. In this figure, the electron density, shown in yellow is mapped onto our structure at the lowest energy point of the conduction band at K (**Fig. SI-16a**), the conduction band minimum at Q (**Fig. SI-16b**), and the valence band maximum at K (**Fig. SI-16c**). The electron density at the K points are purely in the MoSe$_2$ layer at the conduction band and the WSe$_2$ layer in the valence band. At Q however, there is significant electron density in both layers as a consequence of hybridization.

*Analysis of the direct optical transitions near the K-point:*
As explained above, exactly at the K point the states have either pure WSe$_2$ or pure MoSe$_2$ character, therefore optical transitions are suppressed. To find out how rapidly this selection rule is released away from the K-point, we have used the "optics" program that is part of the WIEN2k package[10]. We calculated the joint density of states (JDOS) and the imaginary part of the dielectric function. The ratio of the two gives an average value of the corresponding optical matrix elements weights:

$$J_{DOS}(\omega) = \sum_{\alpha \in \text{unocc}} \sum_{\beta \in \text{occ}} \sum_{\mathbf{k}} \delta(E_{\mathbf{k}\alpha} - E_{\mathbf{k}\beta} - \hbar\omega)$$

$$\text{Im}\varepsilon(\omega) = \frac{e^2}{\pi e m^2 \omega^2} \sum_{\alpha \in \text{unocc}} \sum_{\beta \in \text{occ}} \sum_{\mathbf{k}} \langle \mathbf{k}\alpha|p|\mathbf{k}\beta \rangle^2 \delta(E_{\mathbf{k}\alpha} - E_{\mathbf{k}\beta} - \hbar\omega)$$

In **Fig. SI-17** we show this ratio for two polarizations, in- and out-of-plane, for direct optical transitions between the upper valence band and the two lowest conduction bands. From JDOS one observes that while the formal absorption thresholds for these two bands, which occur at the K point, are at or below 1.2 eV, the matrix elements remain vanishingly small all the way up to ~2 eV.

*Analysis of the indirect K-Q optical transitions:*
The WIEN2K Liner Augmented Plane Wave (LAPW) code can only compute direct optical transitions as described above. Here we present a "poor man's" estimate of the indirect K-Q optical transitions. The code allows decomposition of the wave functions inside each atomic sphere into relativistic spherical harmonics, each of which is characterized by full angular momentum $j$ and its projection $j_z$. The VBM at K is purely a $WSe_2$ state, while the SO-split CBM bands at Q are a mixture of $MoSe_2$ (70-80%) and $WSe_2$ (20-30%) states (See **Fig. 4** in the main text). For instance, there are two relativistic $d$-orbitals with $j_z = ½$, one with $j = 3/2$ and the other with $j = 5/2$. The explicit expressions are, respectively, $\frac{\sqrt{3}|1\downarrow\rangle - \sqrt{2}|0\uparrow\rangle}{\sqrt{5}}$ and $\frac{\sqrt{2}|1\downarrow\rangle + \sqrt{3}|0\uparrow\rangle}{\sqrt{5}}$, where the number before the arrow is the orbital quantum number $L_z$. Non-zero dipole matrix elements occur if (i) $L^{ini} = L^{fin} \pm 1$, (ii) $S_z^{ini} = S_z^{fin}$, and (iii) for circularly polarized light, $j_z^{ini} = j_z^{fin} \pm 1$. Using the calculated projections and explicit expressions for relativistic orbitals we get (semi)quantitative insight into the scale of the allowed decay processes. Our findings are as follows:
   (1) Both low-lying conduction band states at Q have non-zero matrix elements with the VBM at K with roughly equivalent magnitudes.
   (2) For both states at the Q point both positive and negative polarization is possible, but for the lower state (ILE1) the positive polarization dominates over the negative one, and the opposite is true for ILE2.
   (3) The estimated degree of polarization for ILE1 is ~25% (consistent with the experiment), and for ILE2 ~ -6% (consistent in sign, but not in magnitude; the experimental number is close to -35%).
We should emphasize that the above does not include proper phase factors and the plane-wave parts of the LAPW. These factors are computed internally by WIEN2K but are not exported in a convenient manner.

*Band structure details:*
Electronic band structures with energy ranges wider than shown in **Fig. 4** are plotted in this section. **Fig. SI-18** shows the layer character of the bands, **Fig. SI-19** shows the spin direction of the states, and **Fig. SI-20** shows the magnitude of the spins. The energy splittings calculated with density functional theory depend on the choice of exchange-correlation functional. For this work, we used the Generalized-Gradient Approximation (GGA),[11] which is well suited for systems containing surfaces and vacuum regions, such as the $MoSe_2$ / $WSe_2$ bilayer. To check the dependence of our results on the choice of functional, we computed band structures with two other functionals, tabulated in **Table SI-4**. The other functionals are the Local Density

Approximation (LDA)[12], which only depends on the local electron density, and HSE06[13], a hybrid functional which blends some exact exchange with GGA. All functionals find the conduction band minimum is at Q, but the splittings vary significantly between the different functionals.

*Validation of density functional calculations:*
To test the density functional calculations, we optimized the structure of bulk MoSe$_2$ and WSe$_2$ both with and without van der Waals corrections[14,15] as presented in **Table SI-5**. The calculations performed without van der Waals corrections underbind the layers, resulting in unphysically large *c*-lattice constants. Nevertheless, due to inter-layer hybridization, the layers are bound even without van der Waals corrections. Including van der Waals corrections improves the accuracy of the computed lattice parameters. Note that the Mo-Se distance depends only marginally on van der Waals interactions.

**Table SI-4**
Energies of conduction band splittings computed with three functionals using the GGA atomic positions. The columns show the splitting between two lowest conduction band states at Q and K, $\Delta_Q$ and $\Delta_K$, respectively, and the energy difference between the lowest conduction band at K and at Q, $\Delta_{K-Q}$. All functionals find the conduction band minimum at Q. Energies are in meV.

|  | $\Delta_Q$ | $\Delta_K$ | $\Delta_{K-Q}$ |
|---|---|---|---|
| GGA | 44 | 21 | 137 |
| LDA | 48 | 21 | 168 |
| HSE06 | 42 | 46 | 53 |

**Table SI-5**
Optimized structural parameters of bulk MoSe$_2$ and bulk WSe$_2$ computed with DFT as implemented in VASP[16–18]. The generalized-gradient approximation[11] was used. Results are presented both with and without van der Waals (vdW) corrections[14,15]. For each material, the *a* and *c* lattice constant, as well as the metal-Se distance are presented in angstroms (Å). For comparison, experimental results obtained in bulk material at room temperature are included[19,20].

|  | MoSe$_2$ | | | WSe$_2$ | | |
|---|---|---|---|---|---|---|
|  | a | c | Mo-Se | a | c | W-Se |
| Without vdW | 3.32 | 14.98 | 2.54 | 3.32 | 15.15 | 2.54 |
| Including vdW | 3.27 | 12.74 | 2.52 | 3.27 | 12.78 | 2.53 |
| Experiment (RT) | 3.280 | 13.020 |  | 3.282 | 12.960 | 2.526 |

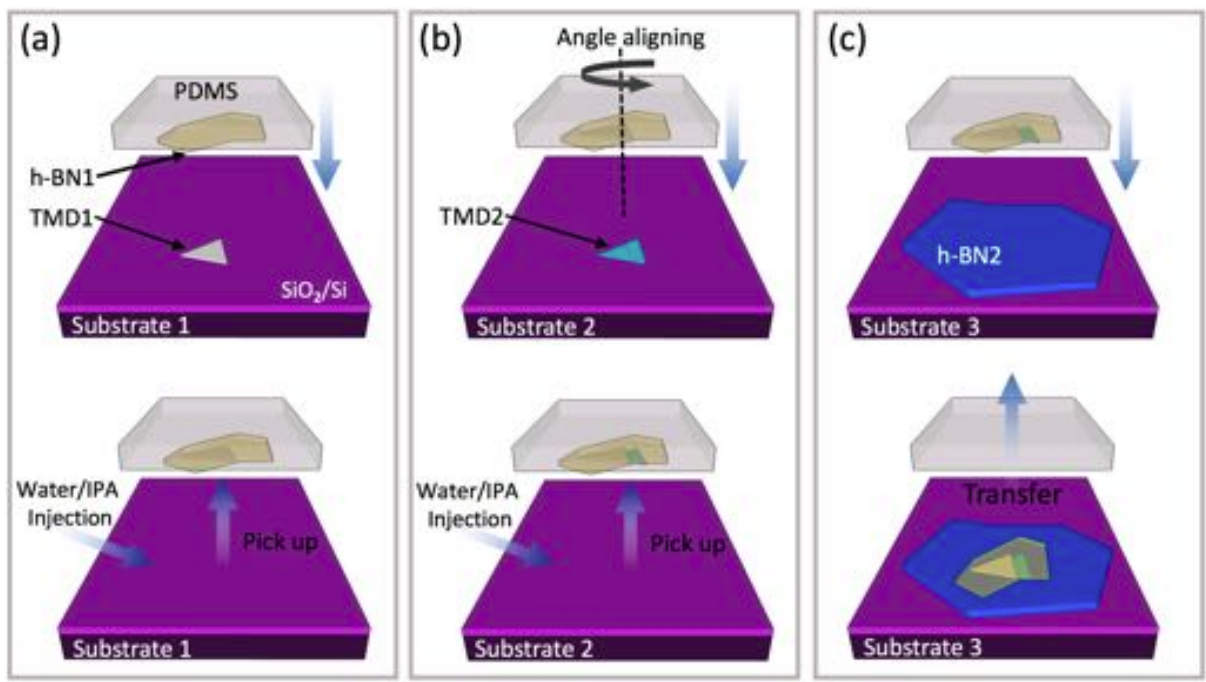

**FIG. SI-1.**
Schematic diagram of the polymer free pick-up transfer method. The steps include (a) pick-up of TMD1 with a PDMS/hBN stack, (b) alignment and pick-up of TMD2 and (c) transfer from PDMS to substrate to complete the heterostructure.

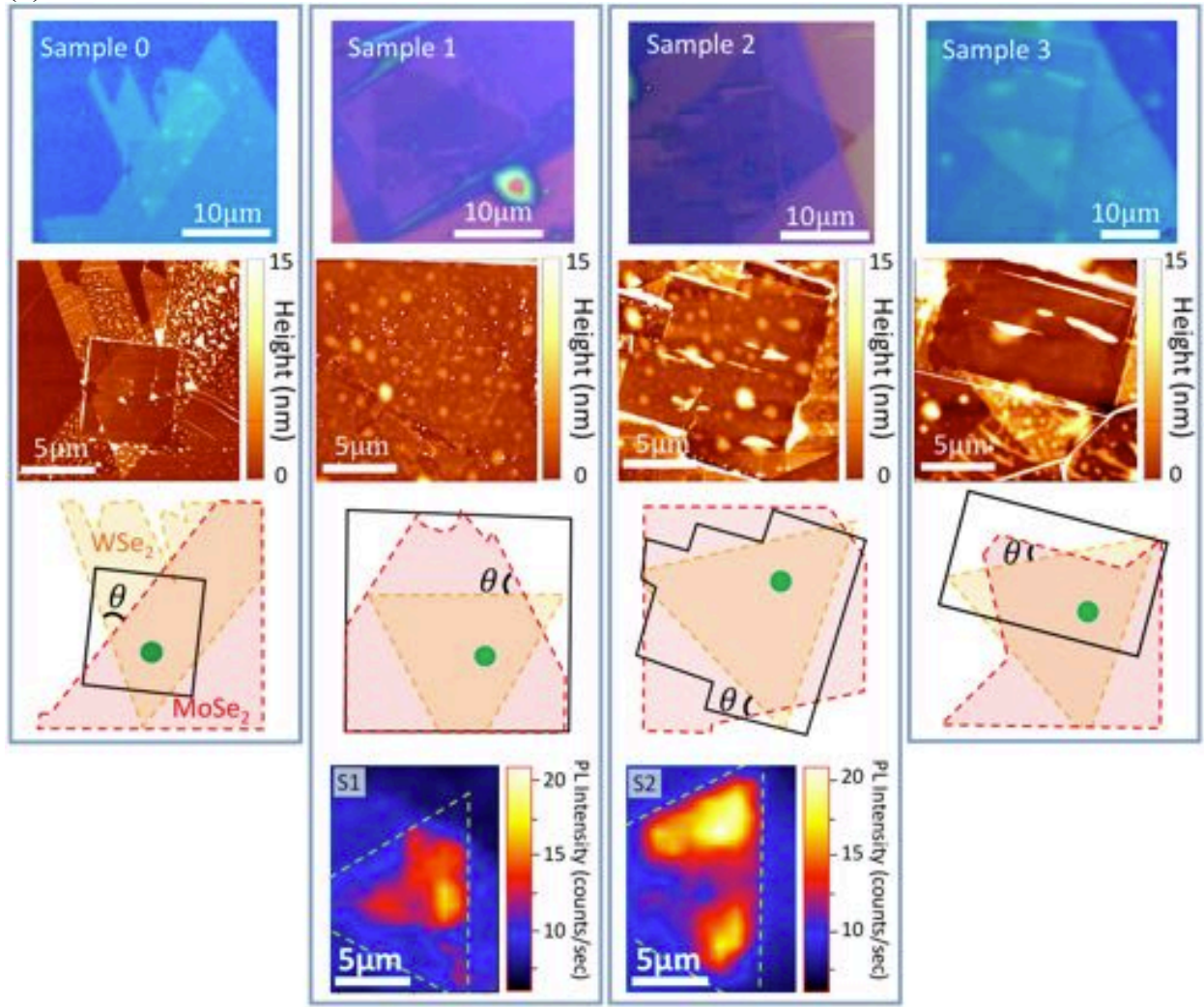

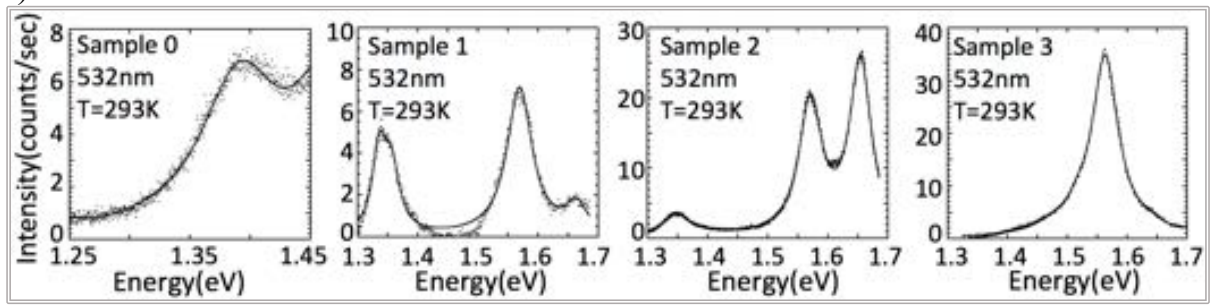

**FIG. SI-2.**
(a) Optical and AFM images of four samples with schematic diagram indicating relative placement of constituents (MoSe$_2$ - red dotted line; WSe$_2$ - blue dotted line). The black solid box indicates the flattened area and green dot shows the laser spot where the (b) room temperature PL measurement were taken. We note the well-aligned samples ($\theta \sim 60°$). S0, S1, and S2 exhibit a clear ILE, whereas no ILE is detected in the misaligned S3 ($\theta \sim 28°$).

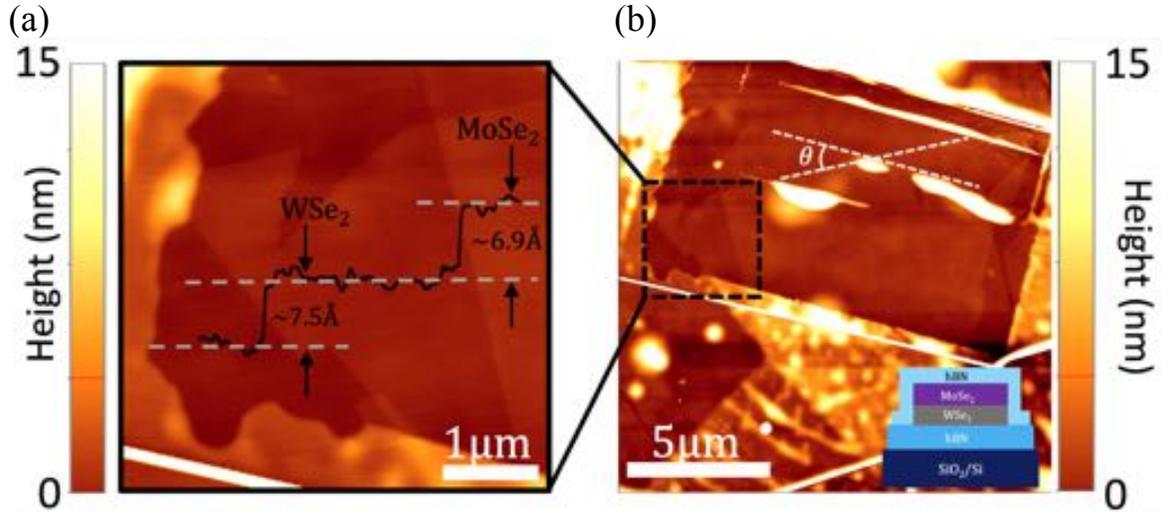

**FIG. SI-3.**
AFM images of S3 at (a) high and (b) low magnification. The inset in (b) is a cross-sectional schematic of the heterostructure.

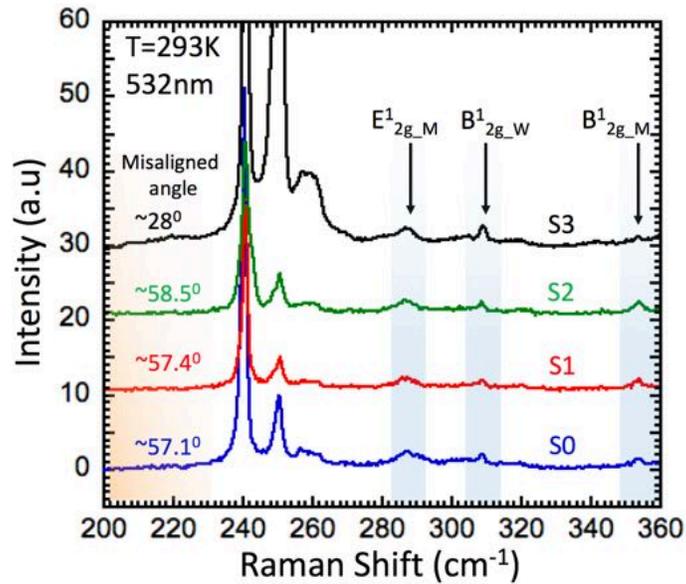

**FIG. SI-4.**
Room temperature Raman spectra acquired with excitation energy of 2.33 eV (532 nm) on representative MoSe$_2$/WSe$_2$ overlap regions in our samples. Spectra are acquired after flattening and have been offset for clarity.

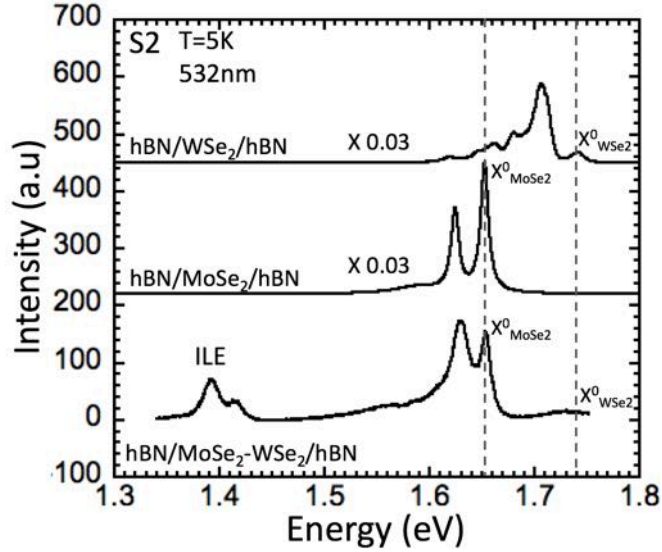

**FIG. SI-5.**
PL spectrum of S2 (bottom) acquired at 5 K with an excitation energy of 2.33 eV (532 nm). For comparison, adjacent MoSe$_2$ (middle spectrum) and WSe$_2$ (top spectrum) monolayers are shown.

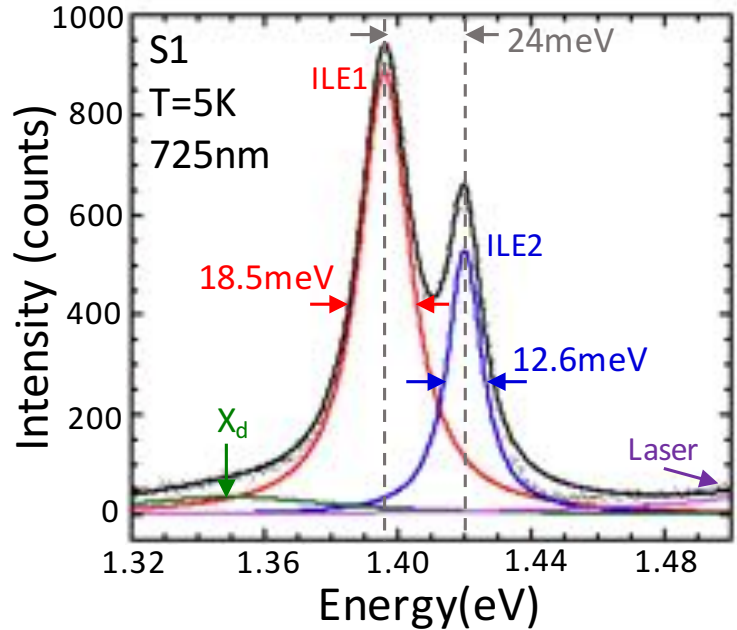

**FIG. SI-6.**
Typical fit of a PL spectrum. Black dots are the data, red and blue solid line are fits to the ILE1 and ILE2 peaks, respectively, and black solid line is the sum of the fitting. The green and purple solid line are a low energy defect peak and the laser tail, respectively.

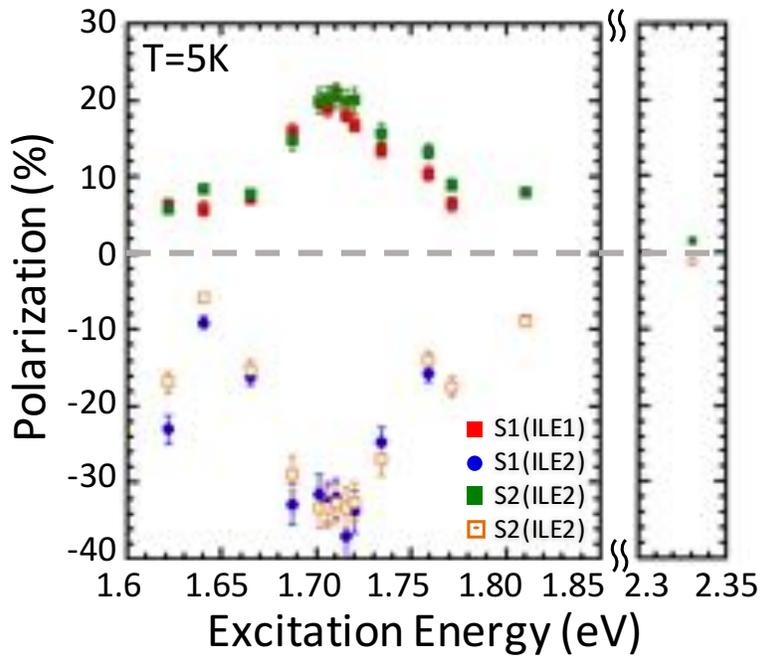

**FIG. SI-7.**
Comparison of polarization as a function of excitation energy for S1 and S2 showing the reproducibility of the polarization of ILE for both samples.

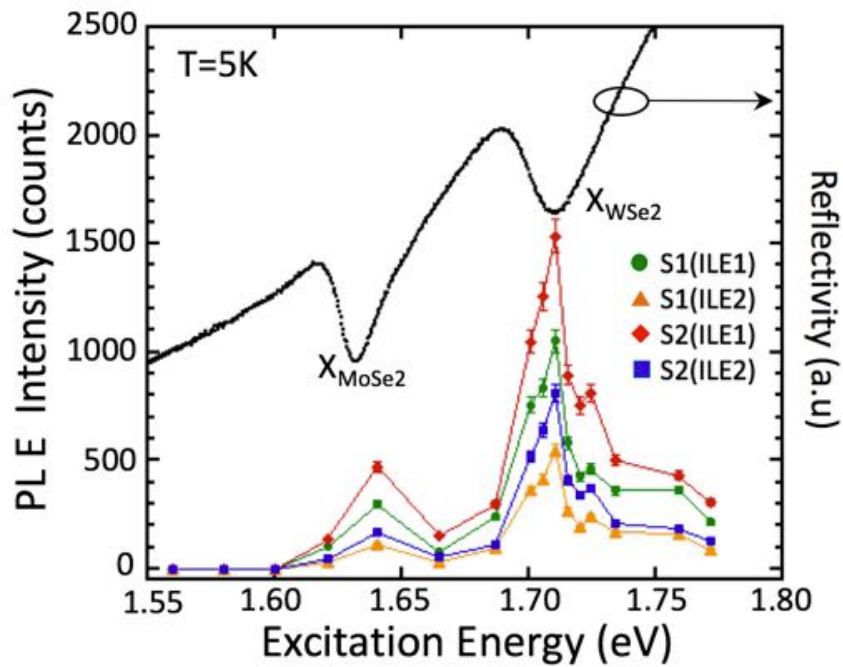

**FIG. SI-8.**
Comparison of PLE for S1 and S2 showing the reproducibility resonance features.

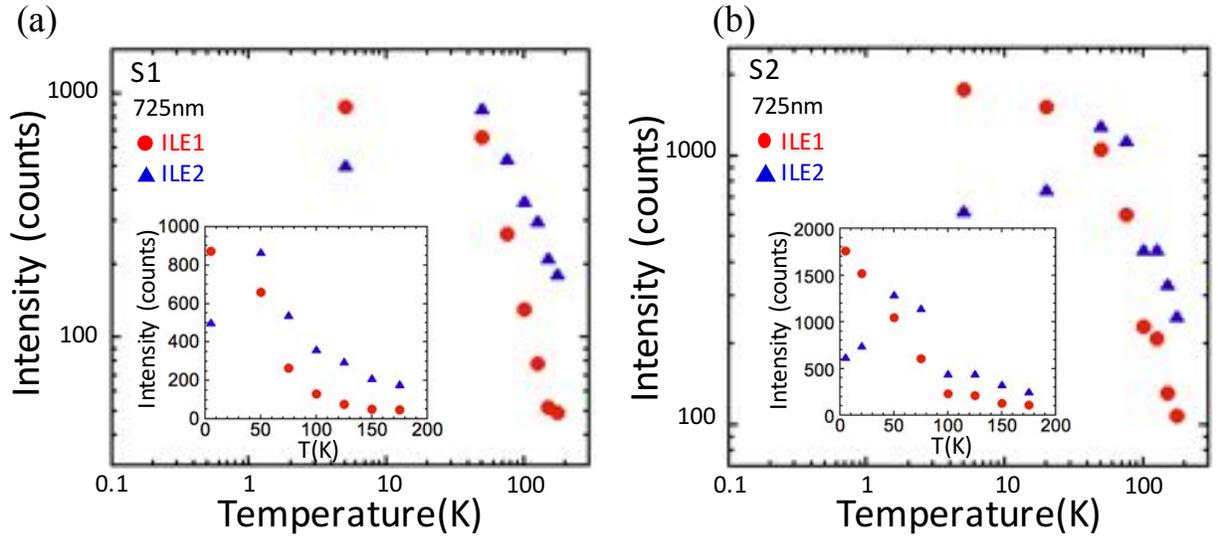

**FIG. SI-9.**
Intensity of ILE1 (red circles) and ILE2 (blue triangles) PL peaks as a function of temperature for (a) S1 and (b) S2. The inset is the same data with temperature on a linear scale. Data were taken with an excitation energy of 1.71 eV (725 nm).

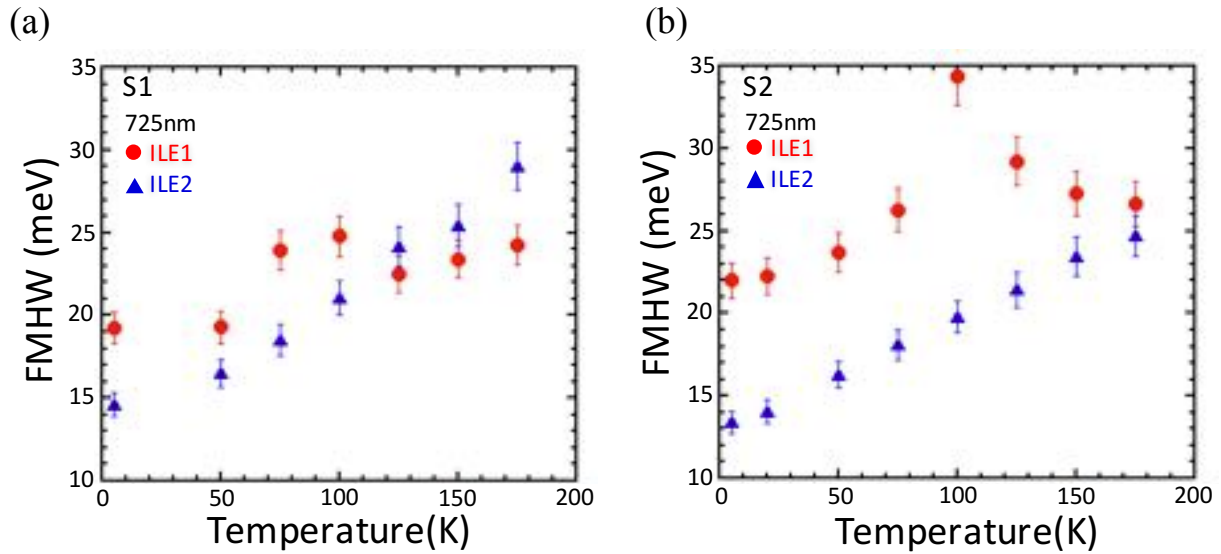

**FIG. SI-10.**
Width of ILE1 (red circles) and ILE2 (blue triangles) PL peaks as a function of temperature for (a) S1 and (b) S2. Data were taken with an excitation energy of 1.71 eV (725 nm).

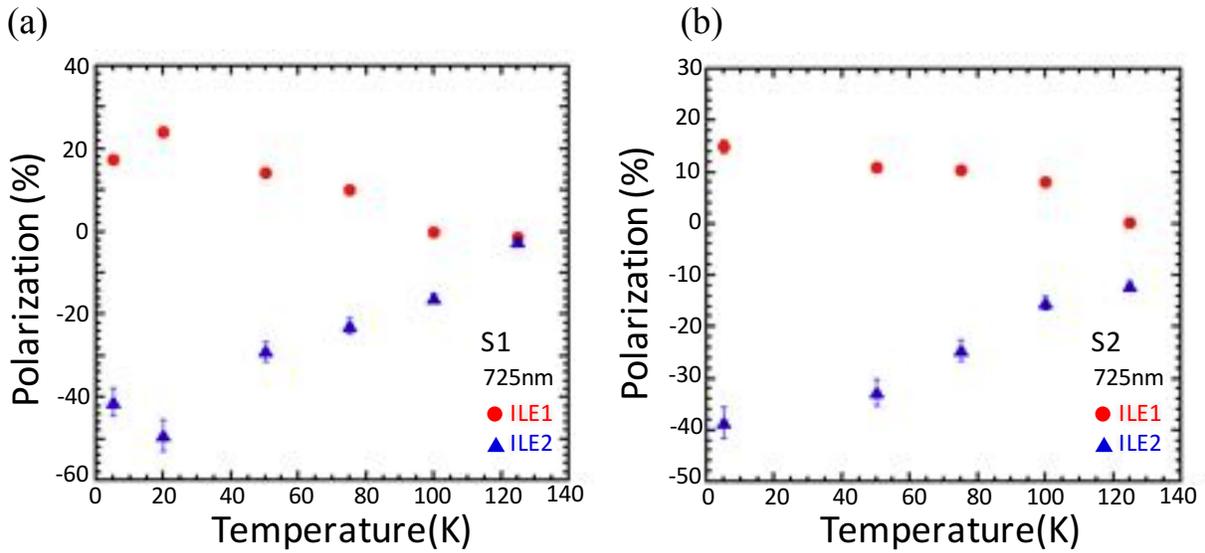

**FIG. SI-11.**
Polarization of ILE1 (red circles) and ILE2 (blue triangles) PL peaks as a function of temperature for (a) S1 and (b) S2. Data were taken with an excitation energy of 1.71 eV (725 nm).

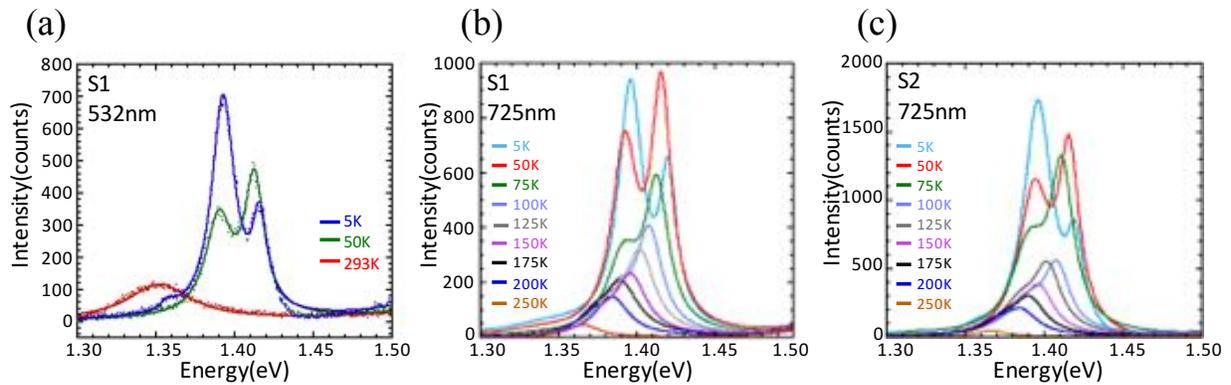

**FIG. SI-12.**
PL spectra of S1 and S2 as a function of temperature. (a) Select PL spectra from S1 measured at 5 K, 50 K and 293 K with an excitation energy of 2.33 eV (532 nm). Temperature dependent (5 K–250 K) photoluminescence from (b) S1 and (c) S2 with an excitation energy of 1.71 eV (725 nm).

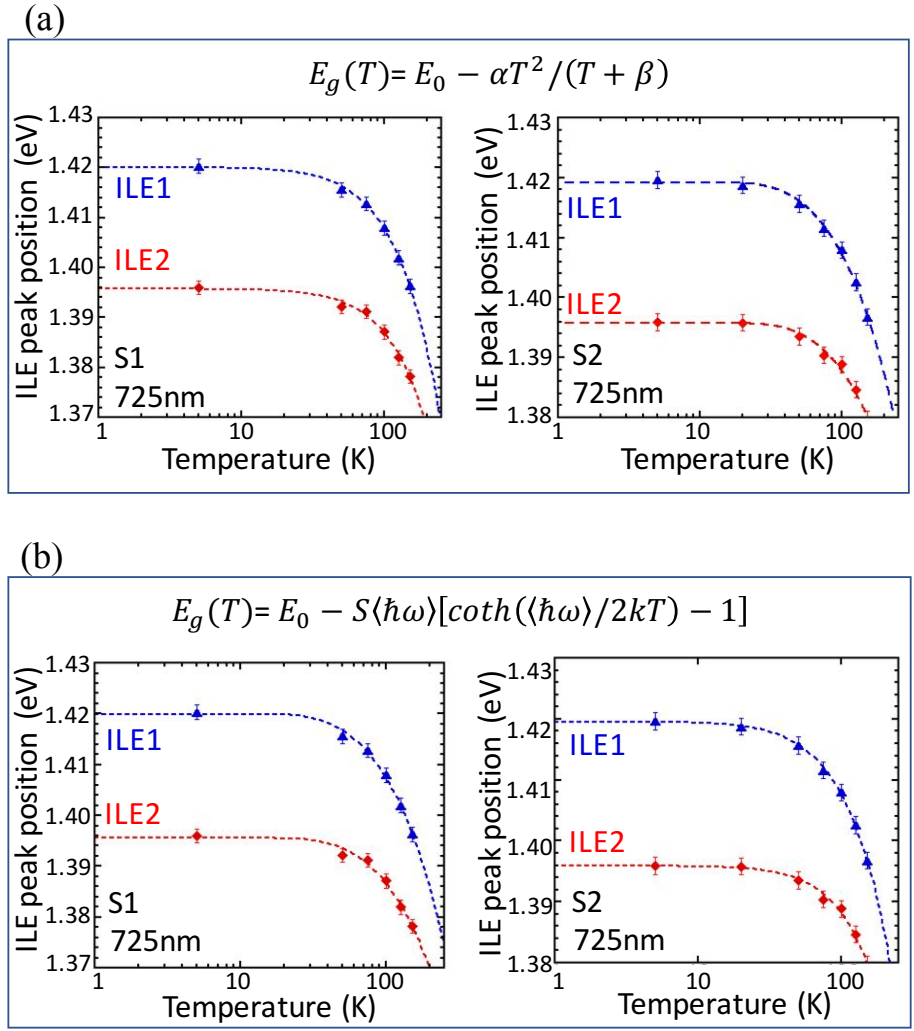

**FIG. SI-13.**
ILE1 (blue) and ILE2 (red) peak position as a function of temperature for S1 (left) and S2 (right). The data are fit (dashed lines) using either the (a) Varshni or (b) O'Donnell formulation. Data are derived from Fig. SI-5.

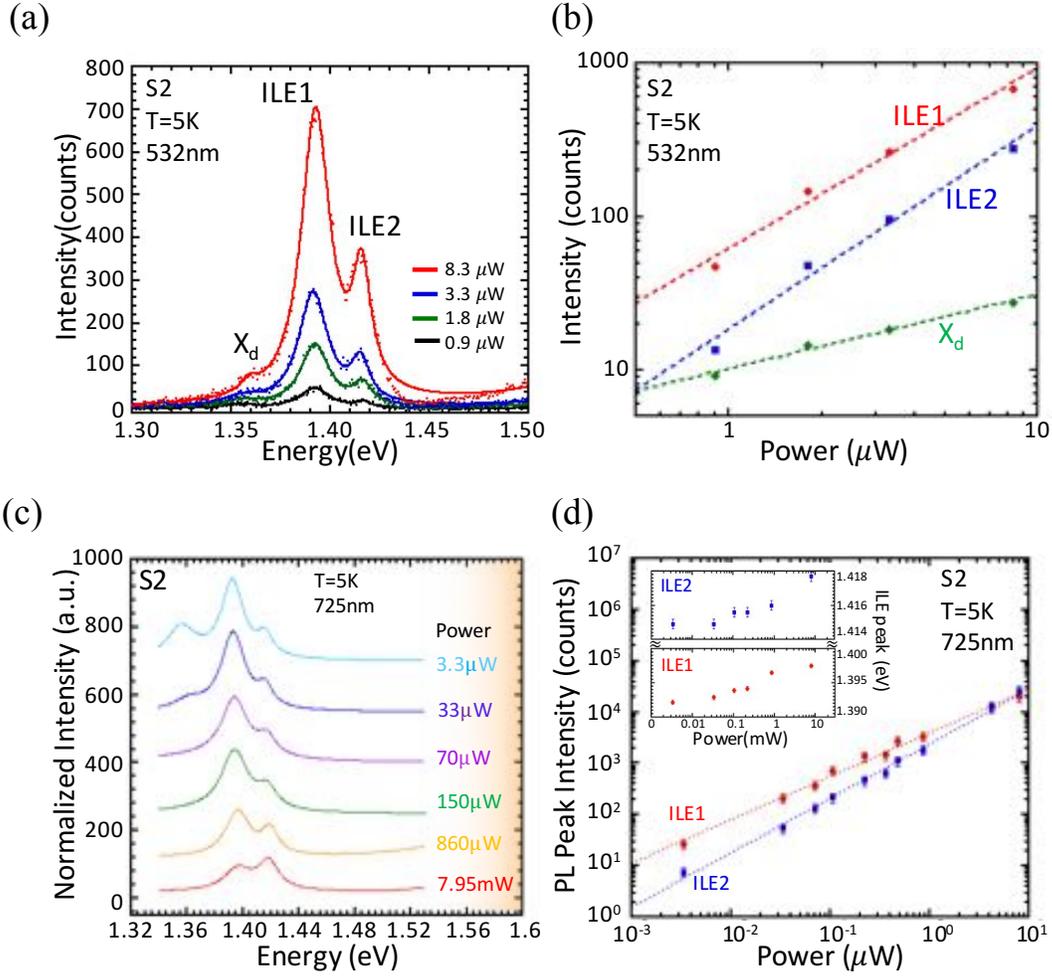

**FIG. SI-14.**
Power dependent measurement of S2 (a,b) at 5 K with excitation energy of 2.33 eV (532nm), and (c) as a function of temperature with an excitation energy of 1.71 eV (725 nm). (d) Summary of ILE peak intensity and peak energy (inset) as a function of power at 5 K with an excitation energy of 1.71 eV (725 nm).

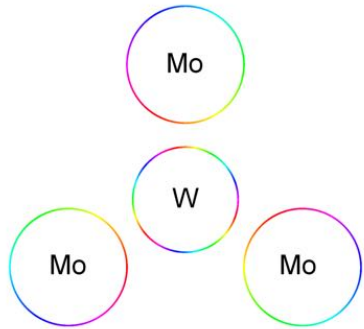

**FIG. SI-15.**
A cartoon illustrating the phases of $L = 2$, $L_z = 1$ wave functions at the K-point. The three outer circles' colors show the phases of the Mo wave functions at the K point of the conduction band. The central circle shows the sum of the three functions decomposed around the W site. One can see that they can only hybridize with L=3 harmonics.

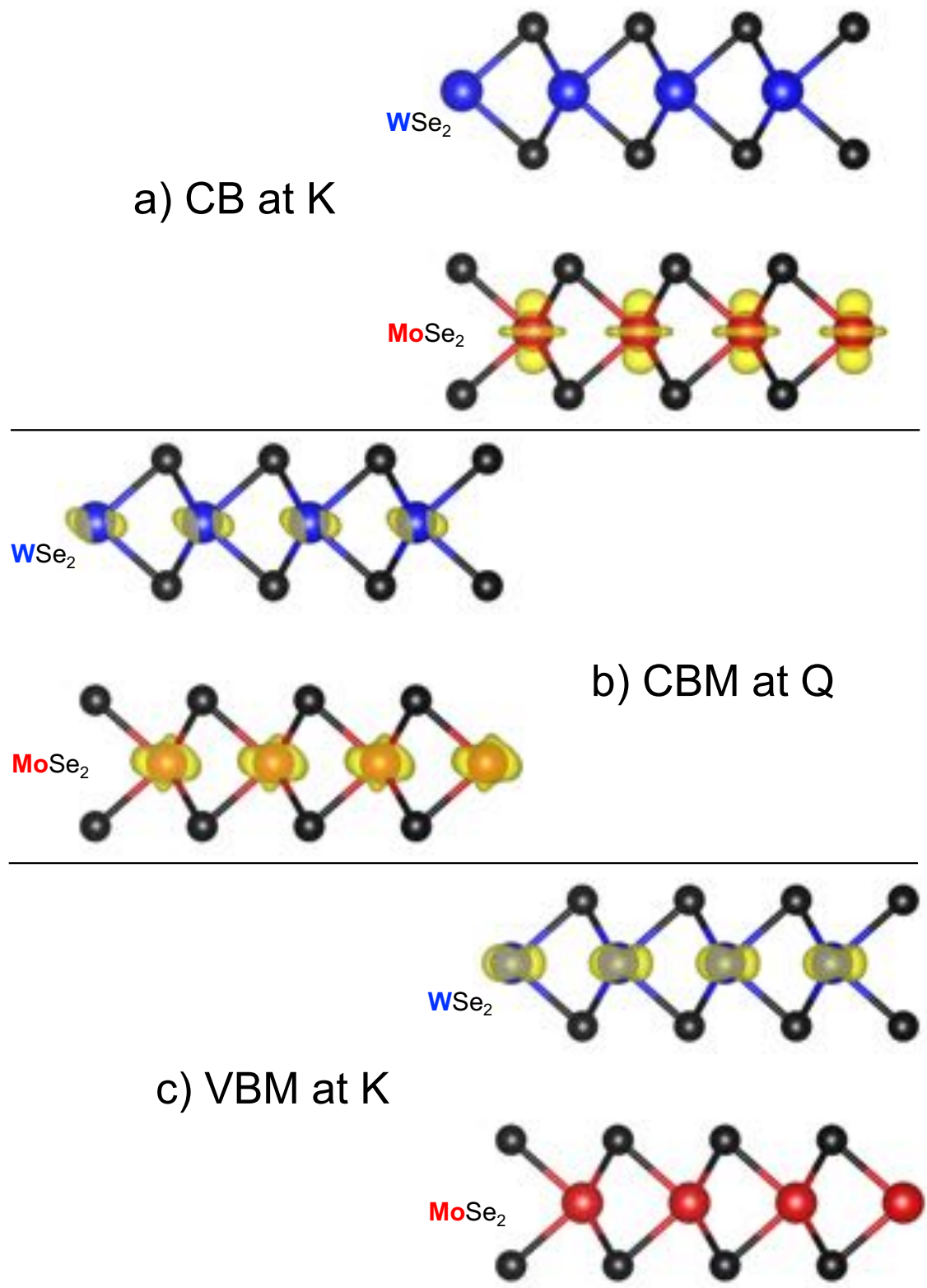

**FIG. SI-16.**
Electron density at three points in the band structure. (a) Lowest conduction band state at K, b) conduction band minimum at Q, c) valence band maximum at K. The Mo are red, W are blue, Se are black, and the electron density isosurfaces are yellow.

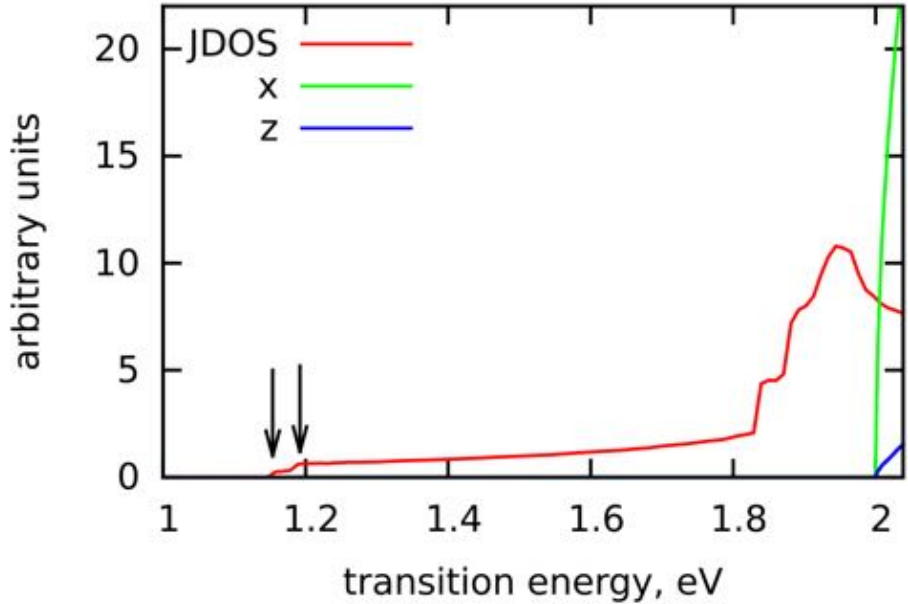

**FIG. SI-17.**
Calculated joint density of states (red) and the imaginary part of the dielectric function for two linear light polarizations, in- and out-of-plane. Arrows indicate thresholds in the joint density of states at the K point, but the optical matrix elements are negligible below ~2 eV. Only direct optical transitions are included in this calculation.

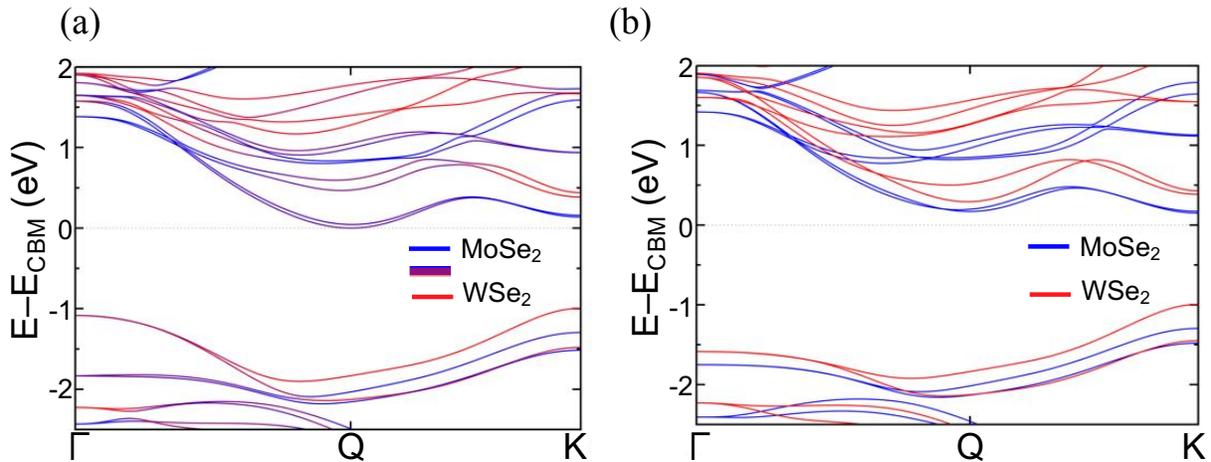

**FIG. SI-18.**
Band structure indicating the layer character of the states. Blue shows states derived from the MoSe$_2$ layer; red from the WSe$_2$ layer and other shades are combinations of these layers. The band structure is plotted for a) the composite heterostructure b) the individual layers with the layers shifted so the VBM at K is at the same energy as in the heterostructure plot.

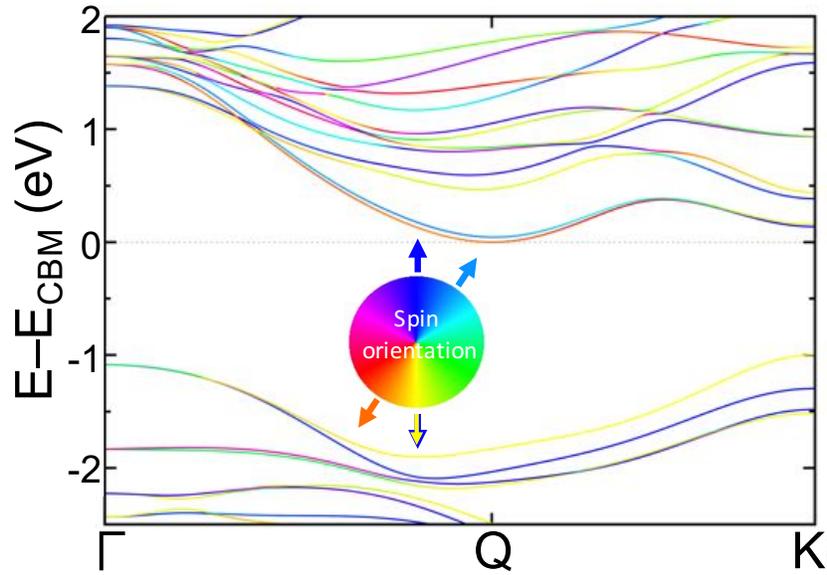

**FIG. SI-19.**
Band structure indicating the spin direction of the states. Blue and yellow represent +z and −z directions, respectively.

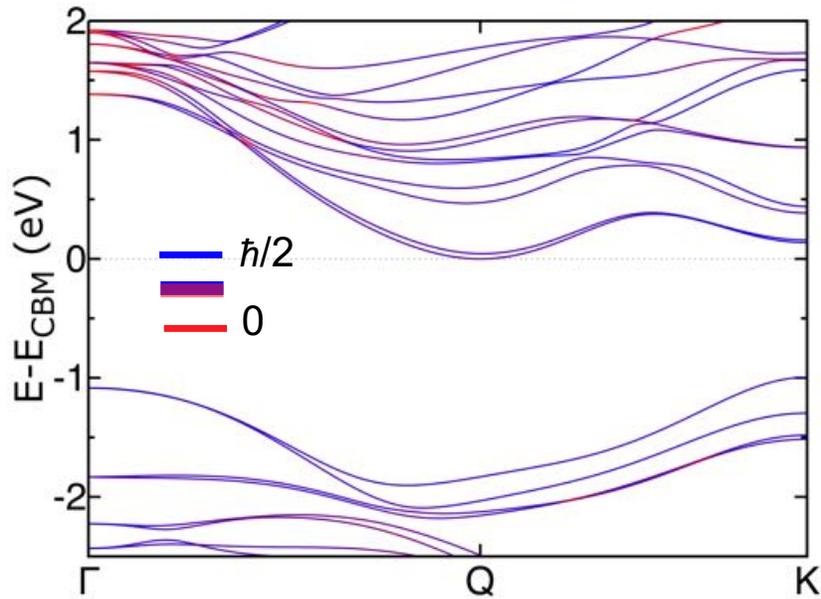

**FIG. SI-20.**
Band structure indicating the magnitude of the spin of the states. Blue indicates $\hbar/2$ and red shows zero magnitude.